# Electrospray techniques for the study of liquid energetics by hyperquenched glass calorimetry

Li-Min Wang, * Steve Borick, # and C. Austen Angell *

\* Department of Chemistry and Biochemistry, Arizona State University, Tempe, AZ 85287-1604

\# Department of Chemistry, Scottsdale Community College, Scottsdale, AZ 85256

## ABSTRACT

We describe an electrospray technique for in situ preparation, for differential scanning calorimetry study, of samples of molecular liquids quenched into the glassy state on extremely short time scales (hyperquenched). We study the case of propylene glycol PG in some detail. Using a fictive temperature method of obtaining the temperature dependence of enthalpy relaxation, we show that the electrospray method yields quenching rates of ~$10^5$ K/s, while the more common method, dropping a sealed pan of sample into liquid nitrogen, yields only 120 K/s. Hyperquenched samples start to relax exothermically far below the glass temperature, at a temperature where the thermal energy permits escape from the shallow traps in which the system becomes localized during hyperquenching. This permits estimation of the trap depths, which are then compared with the activation energy estimated from the fictive temperature of the glass and the relaxation time at the fictive temperature. The trap depth in molar energy units is compared with the "height of the landscape" for PG, the quasi-lattice energy of the liquid based on the enthalpy of vaporization, and the single molecule activation energy for diffusion in hydrogen bonded crystals. The implications for the topography of the energy landscape and the mechanism of its exploration, are considered.

Author to whom correspondence should be addressed: Dr. C. Austen. Angell, caa@asu.edu

## I. INTRODUCTION

In recent studies we [1,2], and others [3,4], have stressed the importance of laboratory hyperquenching strategies in clarifying the physics of glassforming systems. Not only does vitrification on very short time scales help bridge the current gap between computer simulation investigations of supercooled liquids and their experimental counterparts, but it provides glassy materials in much higher states of configurational excitation than have previously been studied in appropriate detail. The properties of high fictive temperature glasses appear to differ in important ways from those of normal glasses, and elucidating these differences will be important in refining our understanding of the structure and energetics of the viscous liquid, and of the vitrification process itself.

The aging process in glasses, both of spin and structural types, has recently been the subject of intense study [5-8]. The studies on glasses performed by molecular dynamics methods [7,8] have been conducted on time scales of tens of ns, utilizing glasses formed on tenths of ns timescales. Current work [9] is extending these time scales to the microsecond domain. By contrast, laboratory studies of aging have been largely conducted on glasses formed on the time scale of minutes, and aged on the time scale of days and weeks. By forming the glasses on very short time scales, however, the aging process can be observed at much lower temperatures [10,11] or, alternatively, on much shorter time scales at higher aging temperatures [12]. By studying the onset of relaxation of the hyperquenched glasses during reheating, the depth of the trap in which the glass was arrested can be determined [2]. Then, by appropriate thermochemical methods (detailed below and elsewhere [1,3]), the temperature at which the system was arrested during the hyperquench can be obtained. Thus hyperquenching studies, which in the past have been used almost exclusively to vitrify systems that normally crystallize, can provide information on the energetics of the "liquid landscape" [13,14] in energy ranges approaching those involved in computer simulation studies. In particular the hyperquenching studies provide access, or at least close approach, to the very interesting "crossover" region that is currently much in focus for the case of fragile liquids [15-18].

Hyperquenching studies, like their computer simulation counterparts, depend for their usefulness on the very large gap in the time scales on which the distinct vibrational and configurational degrees of freedom of a viscous liquid are explored. This separability of degrees of freedom is manifested most directly in the familiar glass transition[15b,19] at which the configurational, but not the vibrational, contribution to the liquid heat capacity drops out. It drops out because its exploration time scale exceeds that of the heat capacity-determining experiment. In effect, the system becomes trapped in one of the immense number of minima on the potential energy hypersurface in configuration space [20-22], and thereafter resembles a crystal (single structure) system, in that only vibrational modes can be excited. For glasses formed at normal rates we can ignore a frequently studied but thermodynamically very weak "secondary relaxation". However between the normal glass transition and the crossover temperature this secondary process is considered to become rapidly stronger [13, 15] and to become the dominant relaxation process at $T_c$. One of the aims of hyperquenching studies is to elucidate this striking inversion.

Although the glass transition is usually observed and recorded during heating experiments, it is more rationally defined during cooling experiments and certainly it is the cooling glass transition that is of interest in hyperquenching experiments. Because the numerical value of the glass transition temperature during rapid cooling must be obtained by studies made after the glass has been formed, it is usually referred to by a distinct name, the "fictive" temperature [23]. In the simulation community, the alternative term "internal" temperature is also used [5-8], though the latter is also defined from the behavior of the glass below $T_g$, rather than estimated by the method of this, and related, papers. Velikov et al [24] have shown that the fictive temperature and the midpoint cooling glass temperature are almost identical in value, at least for single component systems.

The process of maintaining full equilibrium in a liquid (excluding crystallization if this is a possibility) involves the exploration of the potential energy "landscape". So also does the process of approaching equilibrium. For equilibrium to be established, the system point must move between a large enough subset of the basins at the appropriate level on the landscape effectively to have explored them all. What is happening during this exploration process that is needed to establish a true equilibrium between kinetic and potential energies? Clearly the degrees of freedom must communicate. There must be an exchange of energy between the phonon and configuron microstates of the system. At equilibrium, the forward and reverse exchange rates between these microstates must be the same. Otherwise the system is said to be annealing, and its properties are time-dependent. Likewise, when the phonon-configuron exchange, which is responsible for maintaining equilibrium during cooling, becomes too slow, and the system becomes non-ergodic as in glass formation.

The energy of the basin in which the system becomes trapped during continuous cooling is higher the higher the cooling rate. The potential energy of the trapped state can be depicted for different cooling rates, as in Fig. 1, in which the fictive temperatures are also indicated. Note that $T_f$ relates to potential energy while the "real" temperature relates to kinetic energy. When the two temperatures are the same, the system is a liquid. Otherwise it is a glass.

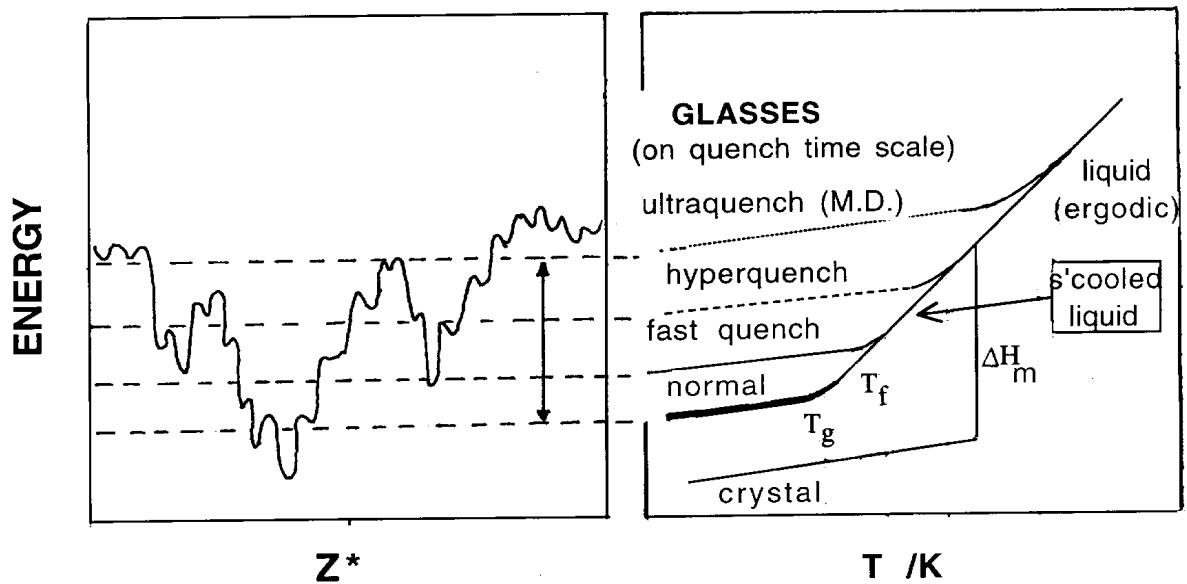

Figure 1. Depiction of the relation between the landscape energy of the basin in which the system is trapped during quenching and the rate of the quench.

While fictive temperatures often can be determined by observations made during cooling, this becomes difficult or impossible when the cooling rate are high. Then the fictive temperature can be assessed by calorimetric measurements performed on the quenched glass during its reheating, as described in detail in recent papers [1,3] which relate back to early studies by Moynihan and co-workers [25]. We will illustrate these features in the present work in which we report on the adaptation of the electrospray technique for differential scanning calorimetry studies of the hyperquenched glassy systems. We adopt the much-studied glassformer 1, 2-propandiol (propylene glycol, PG) as our trial substance and use the data obtained to show how features of the potential energy hypersurface, that are not obvious from ergodic measurements, can be revealed.

## II. EXPERIMENT

Much attention has been given to different methods of quenching liquid systems at high rates, mostly for the purpose vitrifying liquid metals or refining the grain size of crystalline materials [26], but no review will be given here. A technique that has been utilized for the above purpose, but which has also been used to produce finely divided systems (including those containing proteins) for mass spectrometric analysis, is that of electrospraying [27]. In this technique, a metered flow of liquid self-subdivides into very fine droplets, as it emerges from a fine needle orifice, under the mutual repulsion of electrons that are driven onto the liquid surface by a high electrostatic field. It can be refined to the point where even liquid silicon can be obtained as vitreous droplets [28]. It is extremely well suited for producing small samples of liquids sprayed directly into cold pans in the dry environment of the typical differential scanning calorimeter sample housing, and it has therefore been the technique adopted by us for hyperquenching studies.

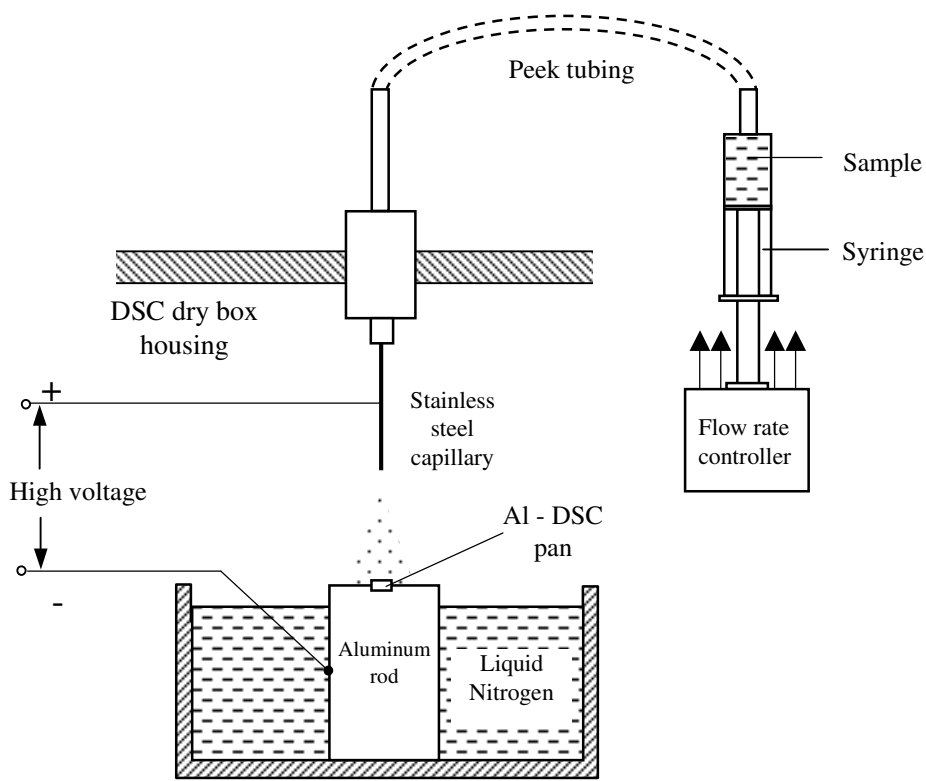

Figure 2. Experimental set-up for electrospraying small samples into cold DSC pan inside dry-box of DSC. Samples are prepared by successive bursts of droplets, allowing time for the first deposit to regain the pan temperature before the next arrives.

The experimental set-up is shown in Fig. 2.  The 1.0 ml capacity syringe, driven by controlled rate stepper motor, supplies liquid through a short length of 0.030″ × 1/16″ peek tubing and adaptor to a 0.004″ × 0.009″ stainless steel needle. The adaptor inserts through the upper perspex plate of the DSC sample head dry box. It inserts directly over an aluminum sample pan that sits on a cold aluminum support rod in contact with liquid nitrogen which is funneled into the mini bath shortly before spraying is to commence. High voltage is applied between the aluminum rod and the syringe needle, the latter contact being simply accomplished by having the syringe needle drop through the splayed end of a multifilament wire lead secured to the dry-box housing.

The voltage, controlled between 6000 and 15, 000 V, is best provided by a power supply designed to have maximum current flows that are non-hazardous on direct contact. Voltages higher than 12,000 V produced corona discharges in the dry box atmosphere of $N_2$. This arcing can be suppressed in a $CO_2$ atmosphere [29] and this will clearly be needed in order to spray high surface tension aqueous systems satisfactorily [30]. The steady flow of liquid, which is an important control parameter [27], was provided by a stepper-motor-driven syringe, the step rate itself being under precise control. The flow rates found best are in the range of 3 – 15 μl/min.

Each liquid sprays optimally under different voltage and flow rate conditions - which must be determined empirically. While we have not yet electrosprayed water effectively, we find that the lower surface tension, but still hydrogen-bonded, liquid propylene glycol, has very satisfactory electrospraying characteristics, and we use it for our initial studies.

To utilize the hyperquenched samples for study of the quenched-in state, and its relaxation to the normal state, we compare the thermal behavior of the hyperquenched sample with a "standard scan" of the same sample. The standard scan [1,24,31] is one in which the heat flow is measured during an upscan at the standard scan rate of 20 K/min after preparing the sample initially by cooling it through the glass transformation range at 20 K/min. For the comparison, the quenched sample is upscanned at the standard rate, and the two scans are superposed using the data at temperatures (i) below that of the relaxation onset of the quenched sample, and (ii) above the glass transformation range, $T > 1.1 T_g$, (where all traces of the sample history disappear) to adjust any slope discrepancies.

## III. RESULTS

In Figure 3 we compare standard scan and quenched sample scans for two cases quenched at two different rates. The upper pair of curves is for a sample that was initially quenched by dropping it, in a hermetically sealed aluminum pan, into a bath of liquid nitrogen, before cold transfer to the DSC head for upscan. The lower pair is for a sample that was electrosprayed into an open DSC pan, as described in the previous section, before cold transfer to the DSC head. The difference between the energies trapped in the two cases is immediately seen, by the much larger area lying between the standard scan and the quenched scan in the case of the electosprayed sample.

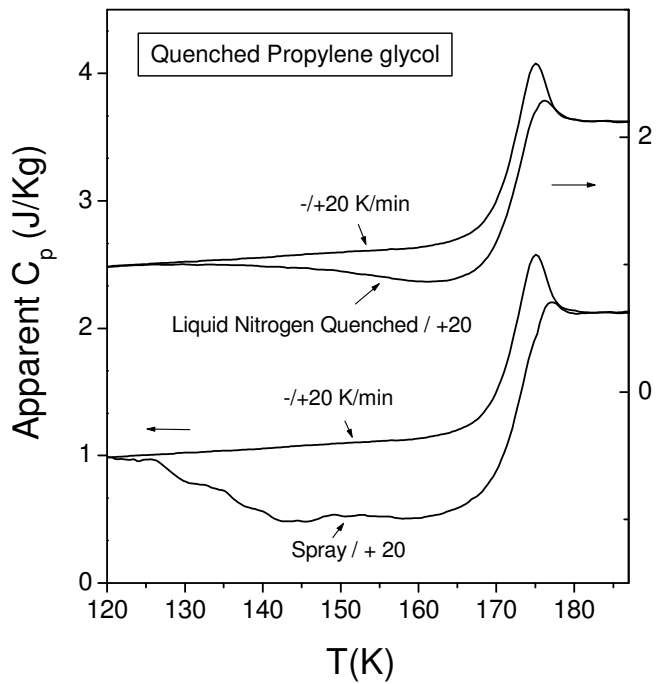

Figure 3. Comparison of the upscan (at standard scan rate) of quenched samples with standard samples for the case of propylene glycol. The upper pair is for a sample, in sealed pan, dropped into liquid nitrogen, while the lower pair is for electrosprayed sample in open pan. The much larger frozen-in enthalpy of the lower case is seen in the much larger area between the two scans for this case.

The effect of changing the cooling rate, under instrument control, between 5 and 40 K/s can be seen in Fig. 4. At these much lower cooling rates the difference between standard and non- standard scans is seen mainly in the immediate vicinity of the glass transition, whereas in the case of higher quench rates the difference is manifested at increasingly lower temperatures. The origin of this distinction will be discussed in the next section.

Fig. 4 is used to obtain the data needed to estimate the quench rate from the data of Fig. 2. The way in which the areas between standard scan and scans at other non-standard cooling rates can be used (see Fig. 4 inset) to obtain the fictive temperatures has been described adequately elsewhere [1,3,31] and need not be repeated in detail here.

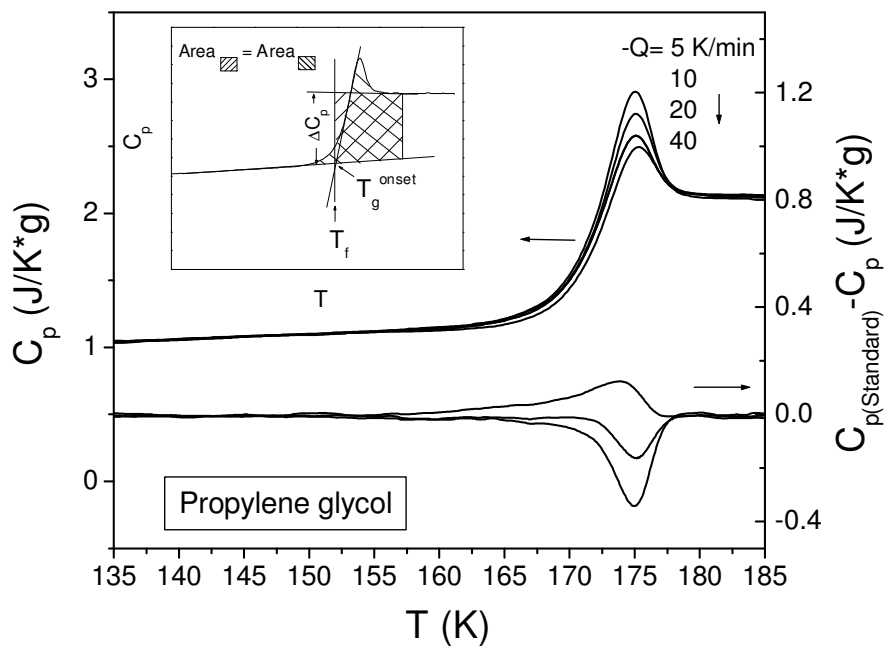

Figure 4. Upscans of samples cooled under instrument control at different rates, including standard rate of 20 K/min, (left hand ordinate) and the differences between standard and non-standard scans (right hand ordinate, lower curves). The insert shows how the fictive temperature $T_f$, used in the Figure 5 plot, is obtained in each case.

The fictive temperatures obtained are plotted in a reduced form in Fig. 5. The slope (and intercept) of this plot both yield the liquid "$m$ fragility" [32] or "steepness index [33], with a precision which is not exceeded by that of any other technique, as recently documented in detail [31]. The value of $m$ obtained from Fig. 5 is 54, whereas that reported on the basis of dielectric relaxation measurements in ref. 34 is 54 and that tabulated in ref. 32 from ac specific heat data [35] is 52.

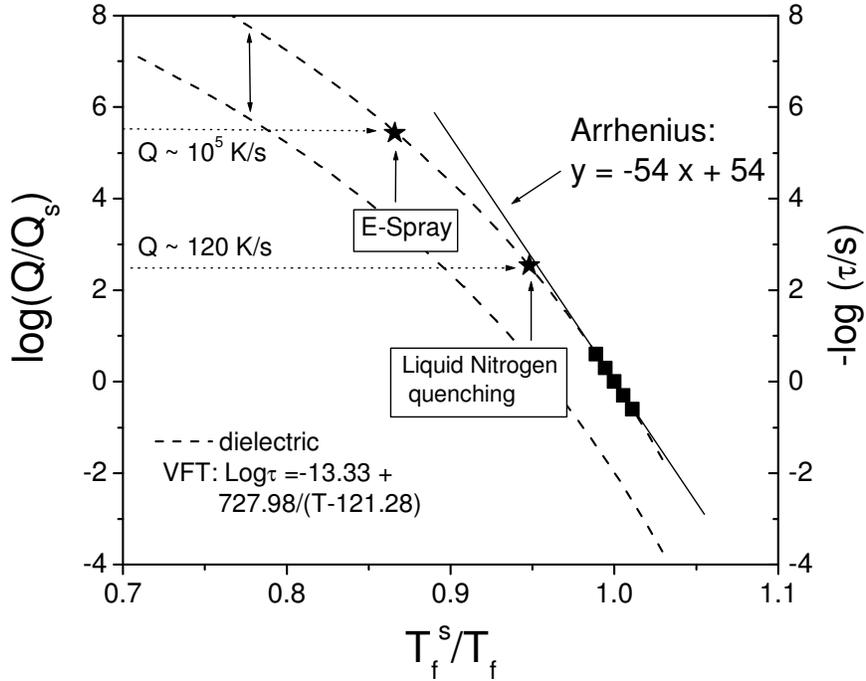

Figure 5. Arrhenius plot of reduced fictive temperature vs. reduced quenched rate, using standard values as scaling parameters. The dash curve is the plot of dielectric relaxation times (from ref. 34), which are matched to the fictive temperature data near $T_f^s$, and used to guide the extrapolation to the reduced fictive temperature of the hyperquenched sample. The value of the reduced quenching rate at this latter reduced fictive temperature yields a quench rate for the hyperquenched sample of $\sim 10^5$ K/s. The parameters of the well-known Vogel-Fulcher-Tammann (VFT) equation for dielectric relaxation times of PG are listed in the legend. The parameters of the Arrhenius equation for the reduced fictive temperature plot give directly the *m*-fragility value for PG, as discussed in ref. 31.

The data of Fig. 5, guided by the form of the dielectric relaxation temperature dependence [34] which is shown as a dashed line in Fig. 5, can now be used to estimate the hyperquenching rate. This is obtained from the intersection with a vertical line representing the scaled fictive temperature obtained for the hyperquenched sample, by the same area integration method. The fictive temperature is found to be 195 K, or 1.15 $T_g$. According to Fig. 5, this indicates a quenching rate of nearly $10^5$ K/s after taking account of the cooling rate of the standard scan, 0.33 K/s. The quench rate we obtain is close to the value estimated for the aerosol droplet method of Mayer [36], applied to the case of di-ols by Mayer and coworkers [12]. Indeed, the temperature at which the (incomplete) upscan of the hyperquenched PG glass in ref. 12 departs from the plots for standard scans [3], is the same as ours to within the uncertainty of determination. In considering the reliability of the Figure 5 estimate of the quenching rate obtained in the electrosprayed sample, it should be borne in mind that the dielectric relaxation times coincide with the ac heat capacity-based relaxation times for PG [35] in the four decade range over which they overlap. (from 4.5 to -0.2 on the -log $\tau$ scale).

By comparison, the liquid nitrogen sealed pan quench is very slow. The fictive temperature obtained from Fig. 2 (upper curves) is only 8.5 K above the standard value of 169 K and this yields, from Fig. 5, a quenching rate of only 120 K/s. Nevertheless, this is a useful quenching rate for some purposes, being 24 times faster than the fastest quench obtainable in the DSC itself (5 K/s).

## IV. DISCUSSION

### A. Calorimetric quantification of initial trap energy level

Using the data of Fig. 3, we can now proceed to estimate the energy, and depth, of the trap in which the sample was arrested during the hyperquench at $10^5$ K/s. The energy of the trap minimum relative to that for the trap in which the standard glass resides, has already been obtained in finding the fictive temperature. This energy gap, given by the area between the hyperquenched and standard scans of Fig. 3, amounts to 1.9 kJ/mole. This is not a useful figure without some relevant scale of energies to which to refer. The most relevant comparisons are with (a) the enthalpy of fusion (b) the enthalpy of exciting the liquid from the glass temperature to the melting point and (c) the enthalpy of exciting the liquid from the Kauzmann temperature to the boiling point. The value of (a) is not available for 1, 2-propandiol, which is extremely difficult to crystallize, but is known for the 1,3 isomer for which it is 12.9 kJ/mol. The value for the 1,2-isomer, when measured, will be similar but smaller: we estimate 12.6 kJ/mol. The values of the enthalpies (b) and (c) can be estimated from the relations

$$\Delta H = \int_{T_g}^{T_m} \Delta C_P dT \qquad (1)$$

and

$$\Delta H = \int_{T_K}^{T_b} \Delta C_P dT \qquad (2)$$

The respective values are 3.69 kJ/mol, and 25 kJ/mol. The latter, which is only approximate for the reasons given below, can be taken as a crude measure of the "height of the energy landscape" for the system, which is conceptually simplest for a constant volume system. The relation between excitation at constant volume (CV) and excitation at constant pressure has (CP) been considered recently [37,38], and it is clear from such considerations that Eq. (2) will overestimate the height of any fixed volume landscape. The overestimate contains three components, none of which will be as important for PG as for most other molecular liquids.

The first component is due to the higher heat capacity at constant pressure over the value at constant volume for which the landscape, and its inherent structures excitation profile, is defined [14]. The extra heat capacity $(C_p - C_v = VT\alpha^2/\kappa_T$ or preferably the configurational part of the difference) must inflate the integrand, but since the expansivity of PG is among the lowest measured for molecular liquids, while the compressibility is normal, the effect will be small.

The second component is due to the changes in the basin shapes that usually occur with level on the landscape, and these changes lead to an excess heat capacity over that due to the configurations alone. The excess is a positive quantity at CP and negative by about the same amount at CV [38-41]. Again, it is expected that this source of difference is quite small for the case of PG which is not a very fragile liquid, as seen above.

The third component comes from uncertainty about the course of $\Delta C_p$ (or $\Delta C_v$) above the melting point. (Of course there is no single melting point, and no boiling point at all, at constant volume, and we use the terms only to indicate approximate points along the excitation profile for the constant volume system - for which a single and unique potential energy hypersurface – or "landscape"- can be defined). $\Delta C_v$ is expected to decrease and eventually to become undefinable as the lifetime of the "structure" becomes comparable with vibrational time scale. By contrast, our estimate has assumed the almost constant value of 72.5 J/mol*K observed in the range $T_g$ - $T_m$.

Taking all these factors into account, we can take the height of the landscape for PG, with a volume fixed at its value at the glass transition, to be about 20 kJ/mol. A precise number is not necessary for our purposes because it is the contrasts with the much smaller trapped-in energy of the hyperquenched glass, and the much larger magnitude of the activation energy for relaxation, on which we wish to focus attention.

Then we can see that, notwithstanding the seven orders of magnitude increase in quenching rate that we have imposed, the energy trapped in the hyperquenched glass in excess of the energy of the standard glass, is rather small relative to the height of the landscape. It is even small relative to the part of the landscape explored between $T_g$ and $T_m$. This reflects the fact that these 7 orders of magnitude in quench rates lie in the regime in which the relaxation time is changing most rapidly with the enthalpy, and accordingly correspond to a relatively small enthalpy change. A change in the quench rate by the same number of orders of magnitude in a computer simulation experiment would involve the system in much larger changes of potential energy because of the relatively larger distance from the Kauzmann temperature in which such a change is being made.

## B. Calorimetric characterization of the initial trap depth

The depth of the basin in which the system is trapped during hyperquenching is a more interesting quantity. It can be estimated in two independent ways from the data obtained in this study. Firstly, it can be estimated from the temperature at which the trapped-in enthalpy starts to be released during the rescan, assuming that the probability of escape from the trap is a Boltzmann function of temperature. Secondly, it can be estimated from the fictive temperature with the help of assumptions about the "true" activation energy for rate processes in viscous liquids. For PG, it is found that both approaches lead to the same trap depth.

In the first approach [2] we use the fact, based on experience with the glass transition, that when an energy change first commences (at the onset glass transition) during a standard scan (20 K/min), the relaxation time for the energy-changing process is 100 s. The energy of the barrier opposing the relaxation can then be obtained by supposing a Boltzmann probability, exp(-$E_{esc}$/RT) per attempt, of escaping from the trap at the temperature $T_{esc}$, and a pre-exponent time constant of $10^{-14}$ s, based on lattice vibrations.

Thus we can write

$$100s = 10^{-14} s \exp\left[\frac{E_{trap}}{RT}\right] \tag{3}$$

from which

$$E_{trap} = 2.303 RT_{esc} \log(10^{16}) = 37 RT_{esc} \tag{4}$$

The trap depth for the present case of hyperquenching, in which $T_{esc}$ is 125 K, is then 39.9 kJ/mole, an energy very much larger than the energy separating the excited glass from the normal glass, as calculated above (1.9 kJ/mol).

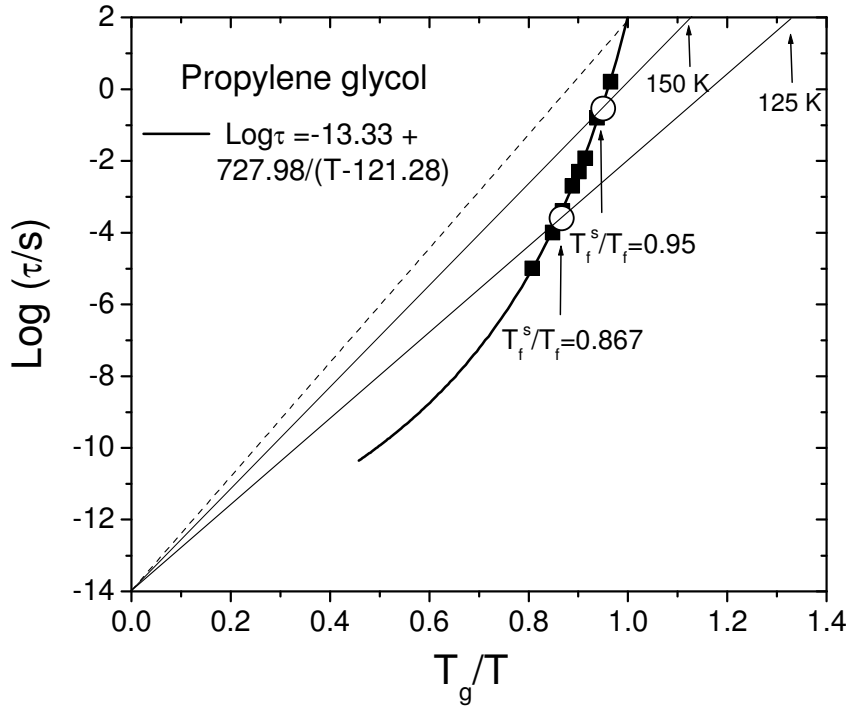

Figure 6. Assessment of the activation energy for relaxation of quenched glasses out of their trap sites, using the scaled fictive temperature, $T_f^s/T_f$. Values for the hyperquenched, and pan-quenched samples of this study, are shown by open circles. Extrapolation of the straight line construction to log $\tau$ = 2, predicts the temperature at which a glass of this fictive temperature will start to relax during warm-up at the standard DSC heating rate of 20 K/min, which are indicated by arrows.

Before discussing this value further, we consider the alternative method of assessing the trap depth. We argue, following Dyre [42] that, at the fictive temperature, $T_f$, located at 1.15 $T_g$, the true activation energy for migration should be given by the construction in which the distorting effect of change of structure with temperature has been removed. Thus, instead of taking the actual slope of the Arrhenius plot for the relaxation time in Fig. 6, the energy barrier opposing the rearrangement of molecules in viscous flow is taken as 2.303R times the slope of the straight line joining the point at $T_f$ on the relaxation time Arrhenius plot, to the lattice vibration time, $10^{-14}$ s. This yields 39.1 kJ/mol for the trap depth, which is very close to that from method 1. Indeed this accord can be seen immediately from the manner in which the extrapolation of the straight line through the fictive temperature to the time $\tau =$ 100 s, yields the temperature 125 K where relaxation begins in Fig. 3.

## C. Interpretation of the trap depth.

The self-consistency of these assessments of the trap depth prompts a series of further questions. For the slower (liquid nitrogen pan immersion) quench (upper plot in Fig.3) the trap depth should be greater. Using the fictive temperature (again obtained from the total energy evolved) to obtain the activation energy, we would predict from Fig. 6 a recovery onset temperature of 150 K. The onset temperature in Fig. 3 is not as well-defined for this case, but there is clearly some weak relaxation below 150 K. This may be due to the spreading out of the relaxation time distribution, which is known to increase with decreasing temperature.

At the glass temperature itself (i.e. at the fictive temperature of the "standard" glass) the activation energy for enthalpy relaxation is obtained as 52 kJ/mol. which is close to the enthalpy of vaporization ($\Delta H_v = 56.7$ kJ/mol at the boiling point of 460.5 K [43]. Before further comment on this correspondence it is helpful to make some comparison with rate processes in other hydrogen-bonded systems, preferably other molecules with two H bonding units per molecule. For instance, the activation energy for diffusion of water molecules in ice [44], which is an Arrhenius process, is 58.5 kJ/mol. This value, which is also obtained for dielectric relaxation, NMR spin-lattice relaxation, and elastic relaxation [44], is greater than the enthalpy of vaporization, 40.66 kJ/mol at the boiling point of water. A more appropriate comparison is with the sublimation energy which is 51 kJ/mol [45,46]. The sublimation energy provides the major component of the derived lattice energy of ice, 56 kJ/mol according to Whalley[45,47].

That the energy fluctuation needed to permit a single particle process like diffusion in a molecular crystal should be comparable to the lattice energy, should cause no surprise. It seems reasonable because the relevant fluctuation involves the expenditure of energy in deforming the lattice sufficiently that a molecular size void is both available and accessible (written as the product of Boltzmann probabilities for void formation and for "jumping" into the void). The manner in which a similar energy fluctuation in an amorphous substance gives rise to diffusion, and thereby to relaxation ($\tau \approx d^2/6D$, where d is the diameter of the rearranging entity, and $D$ is the self-diffusion coefficient) is, on the other hand, not so clear. It is generally believed, after Adam and Gibbs [48] that many particles cooperate in the process and that in this manner the necessity for a molecular sized cavity is avoided. Whatever the detailed mechanism may be, it does not seem surprising to the present authors that, near the glass temperature, the energy fluctuation involved should be comparable to that needed for diffusion in crystals. After all, at the glass temperature, (a) the Arrhenius slopes for processes at constant volume and constant pressure have become essentially the same [38, 49], and (b) the self-diffusion coefficient is smaller than has so far been measured in any *crystalline* substance except silicon [50]. Indeed, the measurement of such a slow diffusion process in solids was only achieved fairly recently [51-54].

Does the relation between the activation energy at $T_g$ and that at the top of the landscape offer any idea of the number of particles (in atomic liquids) or rearrangable sub-units (in complex molecules) that need to cooperate in an elementary relaxation step in liquids near $T_g$? A reasonable estimate might be that obtained by dividing the activation energy at $T_g$ by that at $T_A$, [14,49] (near the normal (1 atm) boiling point) for the case in which the volume remains constant over the whole range. Such data are not available for laboratory systems. However, in the case of mixed LJ the activation energy obtained by the construction of Fig. 6 has increased by nearly a factor of 2 between $T_A$ (where Arrhenius law becomes valid [14]) and $T_C$, (the mode coupling critical temperature or crossover temperature) [14,37]. The number would increase to a value of 5-8 at the temperature extrapolated for a structural relaxation time of 100 s, using the VFT equation provided by ref. 39. This is somewhat greater than the value obtained by Takahara *et al* [55] from application of the Adam Gibbs equation [48], but the latter treatment did not take account of the increase in the *vibrational* heat capacity (over that of the glass), that follows from the change of basin shape with level on the landscape [38,39,40]. Inclusion of that effect in the calculation would lead to a larger cooperative group at $T_g$ than derived by Takahara *et al* [55].

It may be some time before any agreement is reached on this interesting question, but we note here how the construction of Fig. 6 requires that the activation energy at $T_g$ divided by $T_g$ be the same for all liquids. The relaxation at the strong liquid limit presumably occurs by a solid-like single particle process - which would then imply the often-cited idea that the more fragile liquids are more cooperative, i.e. have increasingly large cooperative groups at $T_g$.

**D. Energy landscape considerations**

These considerations have some bearing on the way one must think about energy landscapes. Since the landscape has $3N + 1$ dimensions and a number of minima of the order of exp($N$), there are difficulties in discussing its features directly. It is easier to think first of the landscape for the crystal. Here it is clear that a mechanically stable minimum on the configuration space landscape corresponds to a real space lattice containing a specified number of defects. To move up to a higher level on the landscape it is necessary to create additional defects. In order to change from one basin to another it is evidently necessary to surmount a barrier (the known single particle diffusion activation energy) which is higher than any conceivable level of the landscape, and which approaches the lattice energy in magnitude. "Landscape" is perhaps not the most appropriate description of such an energy topography: A "forest of spikes" would be more accurate for the common 2D representation (cartoon) whereas "rugged honeycomb" might come closer for a 3D representation.

The question to be answered in connection with liquids is the extent to which a similar situation applies to liquid landscapes under conditions where energies and diffusivities approach those of the crystal. At high energies, according to simulations, the landscape broadens out to a high plateau and the most common "basin of attraction" [21] (above which the system "floats") is of very small depth and high anharmonicity [14,55]. Each such minimum can be reached from any other without the need to cross significant barriers. At lower temperatures significant barriers emerge [56]. A question provoked by our observations is whether the liquid landscape evoked by the simulations remains completely distinct from the crystalline case, or whether indeed its fate, at low energies, is to become a "forest of spikes"(in 2D) as in the case of the crystal.

**V. CONCLUDING REMARKS**

By controlled annealing studies such as those recently reported for a hyperquenched mineral glass[3b], the manner in which the hyperquenched glass descends the landscape in successive stages, depending on annealing temperature, can be examined [4]. Extensions of the present work in this direction will be reported in future papers. When a glass has been created suddenly, by increases in pressure, a similar annealing series can be seen. Such stages in the relaxation of HDA (high density amorphous) water, were recently reported, [57] but they were interpreted as evidence for different distinct polyamorphs of water. Extensions of the present work to include samples studied after vitrification at high pressure, may be expected to throw additional light on this interesting question.

**ACKNOWLEDGMENTS**

This work was supported by the NSF under Solid State Chemistry Grant No. DMR 0082535. The authors have profited from stimulating discussions with Andreas Heuer, Jeppe Dyre, Martin Goldstein, and Francesco Sciortino.